\documentclass[aps,prl,groupedaddress,superscriptaddress, twocolumn]{revtex4-2}

\pdfoutput=1

\usepackage[separate-uncertainty,retain-explicit-plus,per-mode=symbol,binary-units]{siunitx}

\usepackage[T1]{fontenc}
\usepackage{hyperref}
\usepackage{xcolor}
\usepackage{graphicx}
\usepackage{array,booktabs}
\usepackage{float}
\usepackage{array,mathtools,dcolumn}
\usepackage{amsmath,amssymb}
\usepackage{multirow} 
\usepackage{lineno}
\usepackage{tabularx}
\usepackage{comment}
\usepackage[normalem]{ulem}
\usepackage[caption=false]{subfig}
\usepackage[version=4]{mhchem}

\usepackage{xspace}
\usepackage[normalem]{ulem}
\usepackage{graphicx}

\DeclareSIUnit\c{\mbox{$c$}}
\DeclareSIUnit\magn{\mbox{$\times$}}
\DeclareSIUnit\min{min}
\DeclareSIUnit\week{week}
\DeclareSIUnit\year{yr}
\DeclareSIUnit\years{years}
\DeclareSIUnit\yr{yr}
\DeclareSIUnit\standard{std}
\DeclareSIUnit\str{sr}
\DeclareSIUnit\ppm{ppm}
\DeclareSIUnit\ppb{ppb}
\DeclareSIUnit\ppt{ppt}
\DeclareSIUnit\pe{PE}
\DeclareSIUnit\spe{SPE}
\DeclareSIUnit\pdm{PDM}
\DeclareSIUnit\ev{events}
\DeclareSIUnit\ct{counts}
\DeclareSIUnit\neutron{\mbox{$n$}}
\DeclareSIUnit\smp{samples}
\DeclareSIUnit\Sample{S}
\DeclareSIUnit\ch{ch}
\DeclareSIUnit\hit{hit}
\DeclareSIUnit\hits{hits}
\DeclareSIUnit\bin{(\mbox{5-PE}~bin)}
\DeclareSIUnit\sgm{\mbox{$\sigma$}}
\DeclareSIUnit\rms{RMS}
\DeclareSIUnit\keVer{\mbox{keV$_{\rm er}$}}
\DeclareSIUnit\keVnr{\mbox{keV$_{\rm nr}$}}
\DeclareSIUnit\eVer{\mbox{eV$_{\rm er}$}}
\DeclareSIUnit\eVnr{\mbox{eV$_{\rm nr}$}}
\DeclareSIUnit\ph{photon}
\DeclareSIUnit\el{\mbox{$e^-$}}
\DeclareSIUnit\pm{\mbox{PMT}}
\DeclareSIUnit\pixel{\mbox{pixel}}
\DeclareSIUnit\inch{''}
\DeclareSIUnit\foot{'}
\DeclareSIUnit\bit{bit}
\DeclareSIUnit\sample{samples}
\DeclareSIUnit\barn{barn}
\DeclareSIUnit\bara{bar}
\DeclareSIUnit\barg{barg}
\DeclareSIUnit\mlardepth{\mbox(meter~of~\LAr~depth)}
\DeclareSIUnit\Curie{Ci}
\DeclareSIUnit\psi{psi}
\DeclareSIUnit\psf{psf}
\DeclareSIUnit\pcf{pcf}
\DeclareSIUnit\parsec{pc}
\DeclareSIUnit\liveday{\mbox{live-days}}
\DeclareSIUnit\days{\mbox{days}}
\DeclareSIUnit\miles{\mbox{miles}}
\DeclareSIUnit\lumens{\mbox{lm}}
\DeclareSIUnit\degreeC{\mbox{$^{\circ}$C}}
\DeclareSIUnit\degreeF{\mbox{$^{\circ}$F}}
\DeclareSIUnit\electron{\mbox{$e^-$}}
\DeclareSIUnit\Euro{\mbox{\euro}}
\DeclareSIUnit\cph{cph}
\DeclareSIUnit\neq{neq}
\DeclareSIUnit\normal{\mbox{N}}
\DeclareSIUnit\nelec{\mbox{$N_e$}}

\newcommand{\DarkPhotonKinMixing}{\ensuremath{\kappa}\xspace}
\newcommand{\SnMixingAngle}{\ensuremath{\left| U_{e4} \right|^2}\xspace}
\newcommand{\DSf}{\mbox{DarkSide-50}\xspace}
\newcommand{\LdmElecCS}{\ensuremath{\bar{\sigma}_e}\xspace}
\newcommand{\DSfDdExposureNew}{\SI{12306(184)}{\kg\day}}
\newcommand{\AxionElecCoupling}{\ensuremath{g_{\mathrm{A}e}}\xspace}
\newcommand{\DM}{\mbox{DM}\xspace}

\newcommand{\NRs}{\mbox{NRs}}

\newcommand{\ERs}{\mbox{ERs}}

\newcommand{\DSfDdLTPostQualCutnumNew}{653.1} 
\newcommand{\DSfs}{\mbox{DS-50}\xspace}
\newcommand{\TPC}{\mbox{TPC}}
\newcommand{\LNGS}{\mbox{LNGS}}
\newcommand{\LAr}{\ce{LAr}}
\newcommand{\GAr}{\ce{GAr}}

\newcommand{\refref}[1]{Ref.~\cite{#1}\xspace}

\newcommand{\SOne}{\mbox{S1}}
\newcommand{\STwo}{\mbox{S2}}

\newcommand{\DSkDriftField}{\SI{200}{\volt\per\centi\meter}}
\newcommand{\DSkExtractionField}{\SI{2.8}{\kilo\volt\per\centi\meter}}
\newcommand{\DSkElectroLuminescenceField}{\SI{4.2}{\kilo\volt\per\centi\meter}}

\newcommand{\PMT}{\mbox{PMT}}
\newcommand{\PMTs}{\mbox{\PMT s}}
\newcommand{\LSVScintillatorMass}{\SI{30}{\tonne}}
\newcommand{\CTFWaterMass}{\SI{1}{\kilo\tonne}}
\newcommand{\reffig}[1]{Fig.~\ref{fig:#1}\xspace}

\newcommand{\refeqn}[1]{Eq.~(\ref{eq:#1})\xspace}

\newcommand{\Eer}{\ensuremath{E_{\mathrm{er}}}\xspace}

\newcommand{\DmLocalDensity}{\ensuremath{\rho_{\mathrm{DM}}}\xspace}
\newcommand{\PhotoelectricCS}{\ensuremath{\sigma_{\mathrm{pe}}}\xspace}
\newcommand{\DmFormFactor}{\ensuremath{F_{\mathrm{DM}}}\xspace}
\newcommand{\IonFormFactornl}{\ensuremath{f^{n \ell}_{\mathrm{ion}}}\xspace}
\newcommand{\LDM}{\mbox{LDM}\xspace}
\newcommand{\ALP}{\mbox{ALP}\xspace}
\newcommand{\ALPs}{\mbox{ALPs}\xspace}
\newcommand{\LdmMass}{\ensuremath{m_\chi}\xspace}
\newcommand{\AlpMass}{\ensuremath{m_{\mathrm{A}}}\xspace}
\newcommand{\DpMass}{\ensuremath{m_{\mathrm{A}^\prime}}\xspace}
\newcommand{\SnMass}{\ensuremath{m_\nu}\xspace}

\newcommand{\APC}{APC, Universit\'e de Paris, CNRS, Astroparticule et Cosmologie, Paris F-75013, France}
\newcommand{\AQLNGS}{INFN Laboratori Nazionali del Gran Sasso, Assergi (AQ) 67100, Italy}
\newcommand{\AQGSSI}{Gran Sasso Science Institute, L'Aquila 67100, Italy}

\newcommand{\AstroCeNT}{AstroCeNT, Nicolaus Copernicus Astronomical Center, 00-614 Warsaw, Poland}
\newcommand{\Augustana}{Physics Department, Augustana University, Sioux Falls, SD 57197, USA}
\newcommand{\Belgorod}{Radiation Physics Laboratory, Belgorod National Research University, Belgorod 308007, Russia}
\newcommand{\BHSU}{School of Natural Sciences, Black Hills State University, Spearfish, SD 57799, USA}

\newcommand{\CAUniPHY}{Physics Department, Universit\`a degli Studi di Cagliari, Cagliari 09042, Italy}
\newcommand{\CAINFN}{INFN Cagliari, Cagliari 09042, Italy}

\newcommand{\CPPM}{Centre de Physique des Particules de Marseille, Aix Marseille Univ, CNRS/IN2P3, CPPM, Marseille, France}
\newcommand{\CTLNS}{INFN Laboratori Nazionali del Sud, Catania 95123, Italy}
\newcommand{\ENUniCEE}{Engineering and Architecture Faculty, Universit\`a di Enna Kore, Enna 94100, Italy}

\newcommand{\FNAL}{Fermi National Accelerator Laboratory, Batavia, IL 60510, USA}

\newcommand{\GEUni}{Physics Department, Universit\`a degli Studi di Genova, Genova 16146, Italy}
\newcommand{\GEINFN}{INFN Genova, Genova 16146, Italy}

\newcommand{\Hawaii}{Department of Physics and Astronomy, University of Hawai'i, Honolulu, HI 96822, USA}
\newcommand{\Houston}{Department of Physics, University of Houston, Houston, TX 77204, USA}
\newcommand{\IHEP}{Institute of High Energy Physics, Beijing 100049, China}

\newcommand{\JINR}{Joint Institute for Nuclear Research, Dubna 141980, Russia}
\newcommand{\Krakow}{M. Smoluchowski Institute of Physics, Jagiellonian University, 30-348 Krakow, Poland}
\newcommand{\Kurchatov}{National Research Centre Kurchatov Institute, Moscow 123182, Russia}

\newcommand{\LNFINFN}{INFN Laboratori Nazionali di Frascati, Frascati 00044, Italy}
\newcommand{\LNHB}{Universit\'e Paris-Saclay, CEA, List, Laboratoire National Henri Becquerel (LNE-LNHB), F-91120 Palaiseau, France}

\newcommand{\LPNHE}{LPNHE, CNRS/IN2P3, Sorbonne Universit\'e, Universit\'e Paris Diderot, Paris 75252, France}
\newcommand{\Manchester}{The University of Manchester, Manchester M13 9PL, United Kingdom}
\newcommand{\MEPhI}{National Research Nuclear University MEPhI, Moscow 115409, Russia}

\newcommand{\MIINFN}{INFN Milano, Milano 20133, Italy}

\newcommand{\MIUni}{Physics Department, Universit\`a degli Studi di Milano, Milano 20133, Italy}
\newcommand{\MSU}{Skobeltsyn Institute of Nuclear Physics, Lomonosov Moscow State University, Moscow 119234, Russia}
\newcommand{\NAINFN}{INFN Napoli, Napoli 80126, Italy}
\newcommand{\NAUniPHY}{Physics Department, Universit\`a degli Studi ``Federico II'' di Napoli, Napoli 80126, Italy}

\newcommand{\Petersburg}{Saint Petersburg Nuclear Physics Institute, Gatchina 188350, Russia}
\newcommand{\PGUniCBB}{Chemistry, Biology and Biotechnology Department, Universit\`a degli Studi di Perugia, Perugia 06123, Italy}
\newcommand{\PGINFN}{INFN Perugia, Perugia 06123, Italy}
\newcommand{\PIINFN}{INFN Pisa, Pisa 56127, Italy}
\newcommand{\PIUniPHY}{Physics Department, Universit\`a degli Studi di Pisa, Pisa 56127, Italy}
\newcommand{\PNNL}{Pacific Northwest National Laboratory, Richland, WA 99352, USA}
\newcommand{\Princeton}{Physics Department, Princeton University, Princeton, NJ 08544, USA}

\newcommand{\RHUL}{Department of Physics, Royal Holloway University of London, Egham TW20 0EX, UK}
\newcommand{\RMTreINFN}{INFN Roma Tre, Roma 00146, Italy}
\newcommand{\RMTreUni}{Mathematics and Physics Department, Universit\`a degli Studi Roma Tre, Roma 00146, Italy}
\newcommand{\RMUnoINFN}{INFN Sezione di Roma, Roma 00185, Italy}
\newcommand{\RMUnoUni}{Physics Department, Sapienza Universit\`a di Roma, Roma 00185, Italy}

\newcommand{\SSUniCHP}{Chemistry and Pharmacy Department, Universit\`a degli Studi di Sassari, Sassari 07100, Italy}

\newcommand{\UCDavis}{Department of Physics, University of California, Davis, CA 95616, USA}
\newcommand{\UCLA}{Physics and Astronomy Department, University of California, Los Angeles, CA 90095, USA}
\newcommand{\UMass}{Amherst Center for Fundamental Interactions and Physics Department, University of Massachusetts, Amherst, MA 01003, USA}

\newcommand{\USP}{Instituto de F\'isica, Universidade de S\~ao Paulo, S\~ao Paulo 05508-090, Brazil}
\newcommand{\VTech}{Virginia Tech, Blacksburg, VA 24061, USA}
\newcommand{\kings}{Physics, Kings College London, Strand, London WC2R 2LS, UK}

\begin{document}

\author{P.~Agnes}\affiliation{\RHUL}
\author{I.F.M.~Albuquerque}\affiliation{\USP}
\author{T.~Alexander}\affiliation{\PNNL}
\author{A.K.~Alton}\affiliation{\Augustana}
\author{M.~Ave}\affiliation{\USP}
\author{H.O.~Back}\affiliation{\PNNL}
\author{G.~Batignani}\affiliation{\PIINFN}\affiliation{\PIUniPHY}
\author{K.~Biery}\affiliation{\FNAL}
\author{V.~Bocci}\affiliation{\RMUnoINFN}
\author{W.M.~Bonivento}\affiliation{\CAINFN}
% ...

\title{Search for dark matter particle interactions with electron final states with DarkSide-50}
\collaboration{The DarkSide Collaboration}\noaffiliation

\begin{abstract}
We present a search for dark matter particles with sub-\si{\GeV\per\c\squared} masses whose interactions have final state electrons using the \DSf experiment's \DSfDdExposureNew\ low-radioactivity liquid argon exposure.
By analyzing the ionization signals, we exclude new parameter space for the dark matter-electron cross section \LdmElecCS, the axioelectric coupling constant \AxionElecCoupling, and the dark photon kinetic mixing parameter \DarkPhotonKinMixing.
We also set the first dark matter direct-detection constraints on the mixing angle \SnMixingAngle for \si{\keV\per\c\squared} sterile neutrinos.
\end{abstract}

\maketitle

The very nature of dark matter (\DM) remains unknown despite cosmological and astronomical observations collecting evidence of its existence over the last several decades~\cite{Faber:1979em,Refregier:2003jl,Clowe:2006hr,Thompson:2015fm,Ade:2016bk}.
Traditionally, \DM particles with masses ranging from a \si{\GeV\per\c\squared} to 
few \si{\TeV\per\c\squared} have been extensively searched for by experiments located in underground laboratories by detecting their interactions with baryonic matter via elastic scattering off atomic nuclei~\cite{PhysRevD.93.122009, PhysRevLett.121.111302, PhysRevLett.118.021303, PhysRevD.98.102006, PhysRevD.100.022004, PhysRevLett.112.041302, PhysRevLett.120.241301, PhysRevD.100.102002} -- usually called nuclear recoils (\NRs).
Heavy \DM can also scatter off electrons, but the energy of such interactions -- called electron recoils (\ERs) -- is suppressed due to the small electron mass.
The lack of concrete evidence of direct \DM detection motivates the search for other candidates and their possible interactions via scattering off, or absorption by, shell electrons, which may subsequently produce sufficiently large ionization signals in the detector~\cite{PhysRevD.85.076007}.

This Letter reports on the analysis of the \SI{\DSfDdLTPostQualCutnumNew}{\liveday} of data 
 collected with the \DSf experiment (\DSfs) to probe \DM interactions in the form of light \DM-electron scattering, absorption of bosonic \DM (axion-like particles and dark photons), and sterile neutrino-electron scattering. 
This analysis uses a more accurate calibration of the detector response~\cite{PhysRevD.104.082005}, improved background modeling and determination of its systematic uncertainties~\cite{DSLongPaper}, and a larger data-set compared to the previous study~\cite{DarkSide:2018bpj}. The same analysis approach  was also applied  to  improve existing  limits on spin-independent WIMP-nucleon interactions  for WIMP masses down to 1.2 GeV/c$^2$ \cite{DSLongPaper} and down to 40 MeV/c$^2$ when including the Migdal effect \cite{DarkSide:2022dhx}. 

\DSfs\ is a dual-phase time projection chamber (\TPC) housed in Italy's INFN Laboratori Nazionali del Gran Sasso (\LNGS).
The active volume consists of low-radioactivity underground liquid argon (\LAr). 
Construction and performance details regarding the \DSfs\ detector are described in \refref{Agnes2015456}. 
Two measurable signals can be observed: the light from scintillation in the liquid (\SOne) and ionization electrons, which are drifted using a \DSkDriftField\ electric field in the \LAr\ volume and extracted by a \DSkExtractionField\ electric field into the gas phase, producing electroluminescence photons (\STwo)
 under acceleration by a \DSkElectroLuminescenceField\ electric field in the \GAr\ volume. 
Two arrays of 19 3-in photomultiplier tubes (\PMTs), one above the anode and one below the cathode, detect photons. 
The \DSfs\ \TPC, enclosed in a stainless steel double-walled, vacuum-insulated cryostat, lies inside a \LSVScintillatorMass\ boron-loaded liquid scintillator veto instrumented by 110 8-in \PMTs\ -- to actively reject neutrons \textit{in situ} -- surrounded by a \CTFWaterMass\ water Cerenkov veto with 80~\PMTs\ -- to actively tag cosmic muons and act as a passive shield against external backgrounds.

The data selection criteria for this analysis aim to identify single-scatter, low-energy events in the form of paired \SOne\ and \STwo\ or \STwo-only pulses uncorrelated to any previous event.
Various quality and selection cuts based on the ratio of \SOne\ and \STwo, \STwo\ signal time profile, and \STwo\ distribution across the \PMT\ arrays are implemented~\cite{DSLongPaper}.
These cuts remove pileup pulses, surface $\alpha$ 
events inducing electrons from the cathode, and spurious trapped ionization electrons released 
up to \SI{20}{\milli\second} after the previous event.
Moreover, only events reconstructed in the fiducial volume are selected, defined by the seven central top \PMTs~\cite{DSLongPaper}.
Veto detector signals are not used in the data selection since \STwo\ triggers are delayed with respect to the veto by the electron drift time in the \TPC.

\begin{figure}[th]
\centering
\includegraphics[width=\columnwidth]{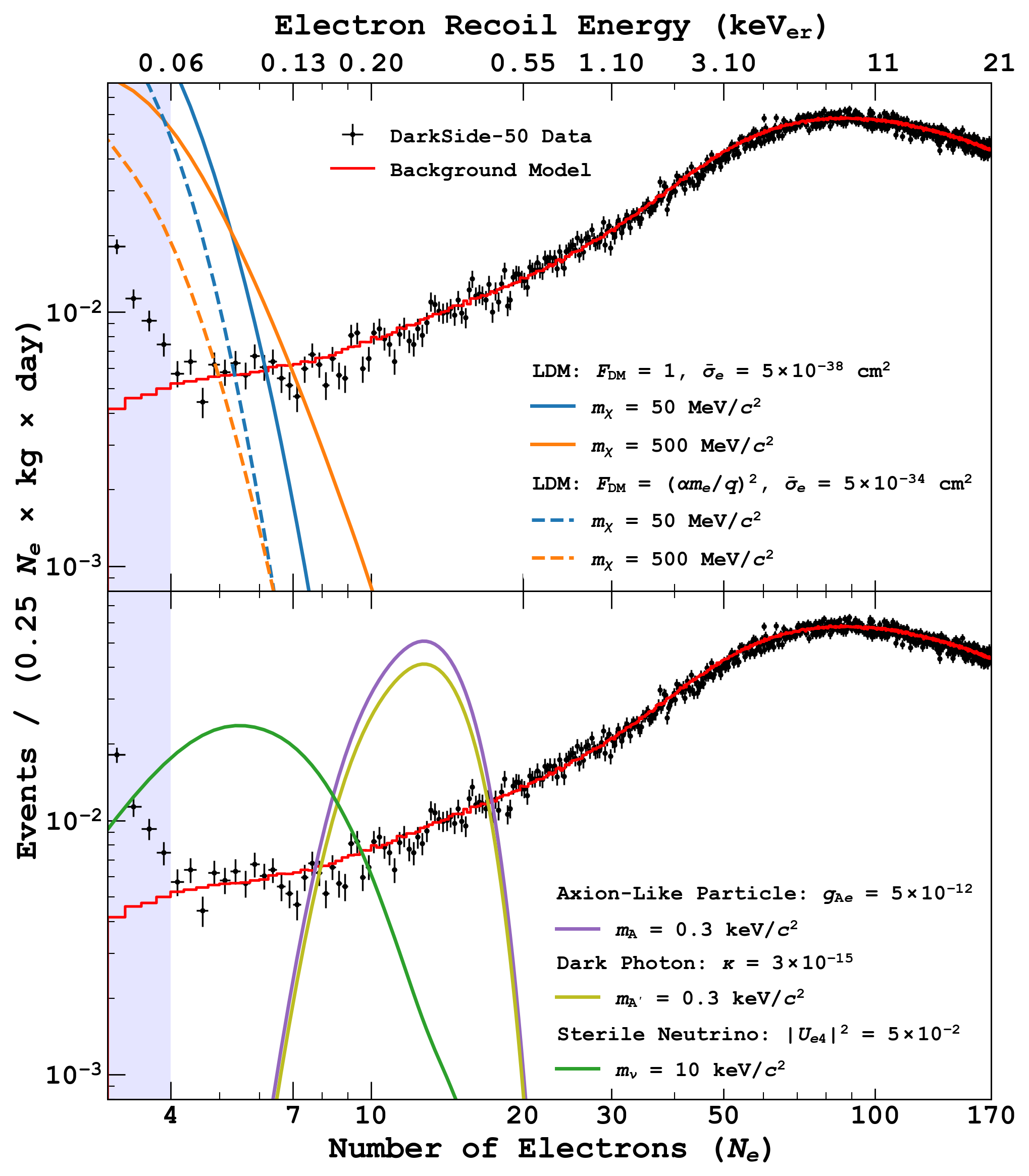}
\caption{Data (black) and background model (red) compared to  expected ionization spectra of illustrative mass values for \DM candidates:  \LDM heavy (blue and orange solid lines) and light (blue and orange dashed lines) mediators, \ALPs (purple), dark photons (yellow), and sterile neutrinos (green). The region shaded blue is below threshold. Details on the background model and its fit can be found in \refref{DSLongPaper}.}
\label{fig:ion_spectra}
\end{figure}

\reffig{ion_spectra} shows the final ionization spectrum obtained after all cuts described in~\cite{DSLongPaper}, resulting in  a fiducial mass of (19.4 $\pm$ 0.3)~kg  and exposure of (12306 $\pm$ 184) kg d.  
This analysis is performed in the energy interval \SIrange[range-units=single]{4}{170}{\el} (\SIrange[range-units=single]{0.06}{21}{\keVer}), up to the range of validity of the detector energy response calibration~\cite{DSLongPaper}. The low threshold is introduced to avoid the region dominated by  spurious electrons.

The background model accounts for the natural radioactivity present in the \LAr\ bulk due to $^{39}$Ar and $^{85}$Kr contamination, and $\gamma$s and X-rays from detector components like the \PMTs, the \TPC\ structure, and stainless-steel cryostat whose specific activities
were determined via a comprehensive material screening campaign. 
For each, the ionization spectra with associated uncertainties were obtained via a detailed Monte Carlo simulation of \DSfs~\cite{DSLongPaper, Agnes_2017}.
The red curve in \reffig{ion_spectra} shows the background model fitted to data.

In this Letter, we search for several \DM candidates using the \DSfs\ data-set and background model.
The candidates are assumed to be non-relativistic and comprise all of the galactic \DM.
While additional local sources for dark matter may be present (e.g. solar production of dark photons~\cite{BlochSearchingDarkAbsorption2017}), we 
set constraints using the interaction rates for the candidate only according to the Standard Halo Model. 
Following the recommendations in \refref{ShawnWhitePaper}, 
we assume a local \DM density (\DmLocalDensity) of 
\SI{0.3}{\GeV \per \c\squared \per \centi\meter\cubed}, 
a standard isothermal Maxwellian velocity distribution ($f(v)$ where $v$ the \DM's velocity) with an escape velocity of \SI{544}{\kilo\meter \per \second}, 
and a local standard of rest velocity of \SI{238}{\kilo\meter\per\second}.
Moreover, the predicted ionization rates per unit mass ($R$) are expressed as a function of the outgoing electron's recoil energy \Eer.
Using the argon ionization response, the spectra are expressed in number of electrons (\si{\nelec}). The ionization response is obtained from the $^{39}$Ar $\beta$-decay sample from an atmospheric argon campaign and from the low-energy $^{37}$Ar peaks. $^{37}$Ar was present in the first few months of \DSfs data and decayed away before the present data-set~\cite{PhysRevD.104.082005}.
The detector response model~\cite{Agnes_2017} is applied to obtain the ionization spectra shown in \reffig{ion_spectra}. 

Fermion or scalar boson light dark matter (\LDM) candidates, with masses below a \si{\GeV\per\c\squared}, can interact with bound electrons via a vector mediator, resulting in the ionization of argon atoms.
The \LDM-electron interaction's dependence on the momentum-transfer $q$ is encapsulated by a dark matter form factor $\DmFormFactor(q)$.
The ionization rate for a \LDM candidate of mass $m_{\chi}$ is parametrized by a reference cross section \LdmElecCS as~\cite{DS50WimpER2018, PhysRevD.85.076007, PhysRevD.96.043017}:
\begin{multline}
\label{eq:ldm_rate}
    \frac{dR}{d \ln{\Eer}} = N_T \frac{\DmLocalDensity}{\LdmMass} \times \frac{\LdmElecCS}{8 \mu^2_{\chi e}} \times \\  
    \sum_{n \ell} \int \left| \IonFormFactornl(k^\prime, q) \right|^2 \, \left| \DmFormFactor(q) \right|^2 \, \eta(v_{\mathrm{min}}) \, q \, dq
\end{multline}
where $N_T$ is the number of target atoms per unit mass, 
$\mu_{\chi e}$ is the \DM-electron reduced mass, 
$\IonFormFactornl(k^\prime, q)$ is the ionization form factor modeling the effects of the bound-electron in the $(n, \ell)$ shell and outgoing final state,
$k^\prime = \sqrt{2 m_e \Eer}$ is the electron recoil momentum, 
and $\eta(v_{\mathrm{min}}) = \int \frac{1}{v} f(v) \Theta(v - v_{\mathrm{min}}) \, dv$ 
is the inverse mean speed function that encodes the \DM velocity profile for the minimum \DM velocity ($v_{\mathrm{min}}$) required to eject an electron with $E_{\mathrm{er}}$ given $q$~\cite{DS50WimpER2018}.
Two benchmark interaction models are considered: a heavy mediator (mass $\gg \alpha m_e$) with $\DmFormFactor = 1$ and a light mediator (mass $\ll \alpha m_e$) with $\DmFormFactor = (\alpha m_e / q)^2$, where $\alpha$ is the fine structure constant and $m_e$ is the electron mass.

\begin{figure*}[htpb]
\centering
\includegraphics[width=\textwidth]{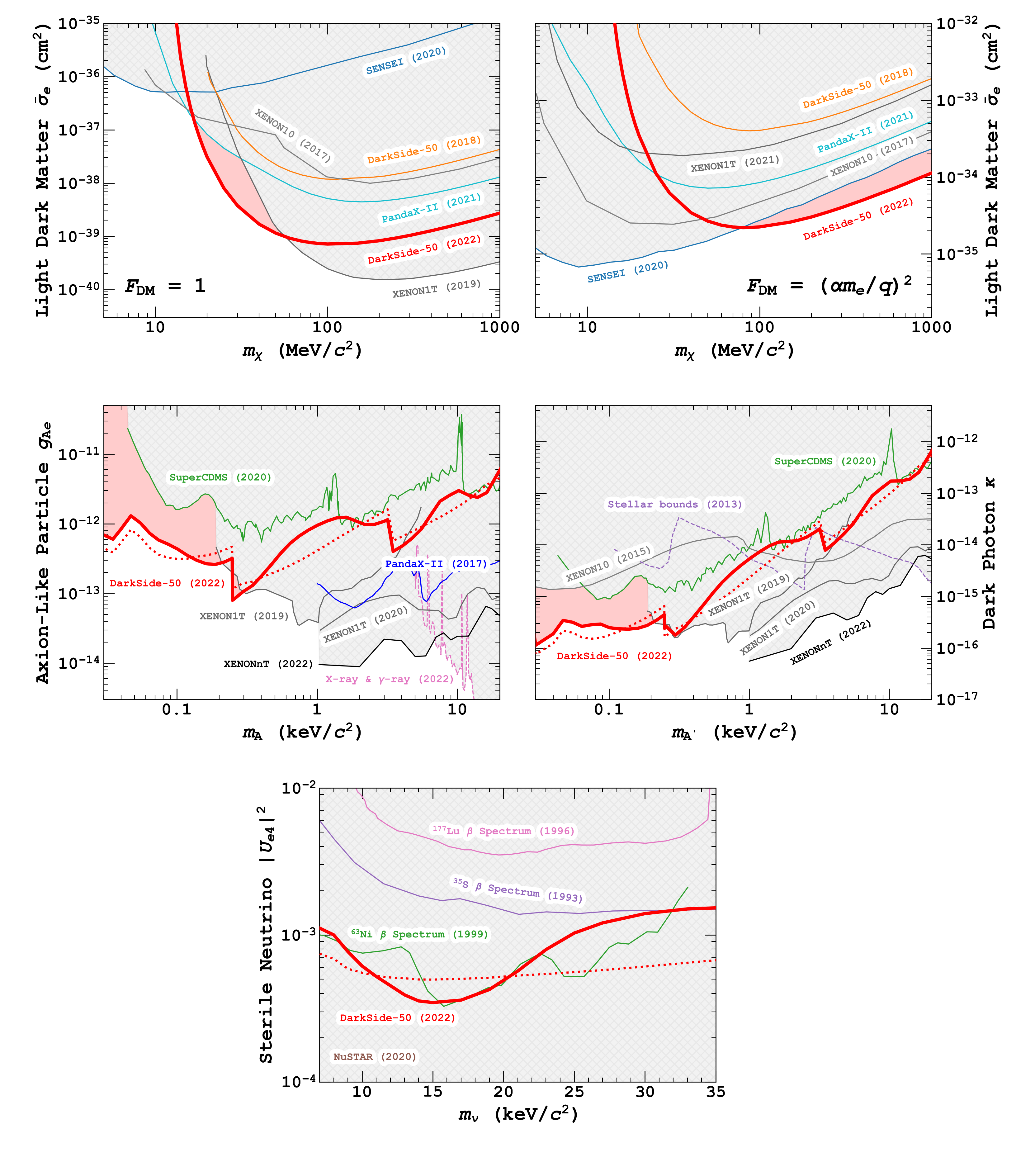}
\caption{Exclusion limits at 90\%~C.L. on \DM particle interactions with electron final states. The x-axis shows the mass of the candidate while the y-axis shows the model parameter. 
Limits set by this work are shown as solid red lines while the $-1\sigma$ expected limits are dotted red lines, and newly excluded parameter space is shaded red.
Limits from laboratory experiments, shown as solid lines, are set by
$\beta$ spectrum analyses~\cite{HolzschuhSearchHeavyNeutrinos1999, MortaraEvidence17KeV1993, SchonertExperimentalLimitsHeavy1996},
\DSfs~\cite{DS50WimpER2018}, 
PandaX-II~\cite{PhysRevLett.126.211803, PhysRevLett.119.181806}, 
SENSEI~\cite{Sensei2020},
SuperCDMS Soudan~\cite{AralisConstraintsDarkPhotons2020},
XENON10~\cite{PhysRevD.96.043017, AnDirectDetectionConstraints2015}, 
and XENON1T~\cite{Xenon1tLightDM, AprileEmissionSingleFew2021,
AprileExcessElectronicRecoil2020}, 
with previously excluded parameter space shaded gray.
Astrophysical constraints (dashed lines) are set by \refref{ViauxNeutrinoAxionBounds2013, FerreiraDirectDetectionExperiments2022, AnNewStellarConstraints2013, RoachNuSTARTestsSterileneutrino2020}.
For sterile neutrinos, the limits set by NuSTAR~\cite{RoachNuSTARTestsSterileneutrino2020} extend downwards to $\SnMixingAngle = 10^{-13}$ at \SI{20}{\keV\per\c\squared}.
%
%
%
%
All limits using the Standard Halo Model are scaled to a local dark matter density \DmLocalDensity of \SI{0.3}{\GeV\per\c\squared\per\cm\cubed}.}
\label{fig:limits}
\end{figure*}

\DSfs is also sensitive to pseudo-scalar \DM such as axion-like particles (\ALPs)~\cite{PhysRevD.90.062009, PhysRevLett.118.261301, PhysRevLett.119.181806, ARISAKA201359} or vector-boson \DM like dark photons~\cite{PhysRevD.78.115012} through absorption by argon shell electrons.
Absorption of either candidate would result in a monoenergetic signal at the particle's rest mass, \AlpMass for an \ALP or \DpMass for a dark photon.
The absorption rate per unit mass of galactic ALPs depends on the axion-electron coupling strength \AxionElecCoupling~\cite{ARISAKA201359, PhysRevD.78.115012}, 
\begin{equation}
    \label{eq:alp_rate}
    R = N_T \frac{\DmLocalDensity}{\AlpMass} \times 
    \frac{3 \AlpMass^2 \AxionElecCoupling^2}{16 \pi \alpha m_e^2} \PhotoelectricCS(\AlpMass c^2) c
\end{equation}
while that of dark photons depends on the strength of the kinetic mixing $\kappa$ between the photon and dark photon~\cite{PhysRevD.78.115012},
\begin{equation}
    \label{eq:dp_rate}
    R = N_T \frac{\DmLocalDensity}{\DpMass} \times 
    \kappa^2 \PhotoelectricCS(\DpMass c^2) c
\end{equation}
where \PhotoelectricCS is argon's photoelectric cross section~\cite{HenkeXRayInteractionsPhotoabsorption1993} evaluated at the particle's rest energy.


A sterile neutrino $\nu_s$ with a mass between \SI{7}{\keV\per\c\squared} and \SI{36}{\keV\per\c\squared} can be a viable \DM candidate~\cite{Leo_2017, 10.1093/mnrasl/slu034}.
Sterile neutrinos interact via $\nu_s + e \rightarrow \nu_e + e$ (and $\bar{\nu}_s + e \rightarrow \bar{\nu}_e + e$)~\cite{Campos_2016}, where a $\nu_s$ mixing with an active state -- parameterized by the mixing angle \SnMixingAngle\ -- inelastically scatters off a bound electron in the detector. 
The ionization rate is governed by the cross section $\sigma_{n \ell}$ between $\nu_s$ and an electron in a given orbital $(n, \ell)$~\cite{Campos_2016}:
\begin{multline}
\label{eq:sterile_nu_rate}
    \frac{dR}{d \Eer} = N_T \frac{\DmLocalDensity}{\SnMass} \times \\
    \sum_{n \ell} 2(2\ell+1)  
    \int \frac{d\sigma_{n \ell}}{d \Eer}(v, \SnMass, \SnMixingAngle) ~ f(v) \, v \,dv
\end{multline}
where $m_{\nu}$ is the sterile neutrino's mass.
We note that \refeqn{sterile_nu_rate} does not include the effects of the ion on the outgoing electron, unlike the treatment of LDM-electron scattering in \refeqn{ldm_rate}.
Additionally, we note that the scattering rates calculated in~\refref{Campos_2016} fail to evaluate the \DM velocity distribution in the lab frame; however, this is corrected in our work.

This analysis employs a binned Profile Likelihood Ratio (PLR) approach~\cite{Cowan2011, Moneta:2011bQ} to determine exclusion limits for each \DM candidate.
The PLR includes a set of nuisance parameters representing the nominal values and uncertainties on the exposure, materials screening, theoretical energy spectra shape, and ionization energy scale. 
Correlations among the different components are encoded in the likelihood definition. 
See \refref{DSLongPaper} for a complete description of the above.  We also verified that the expected limits have negligible dependence on ER fluctuations by changing the Gaussian model described in \refref{DSLongPaper} to a binomial one. 

\reffig{limits} shows the 90\%~C.L.~exclusion limits placed on each candidate via the PLR method. 
The observed limits (solid red lines) are shown alongside the expected limits at $-1\sigma$ (dotted red lines) for \ALPs, dark photons, and sterile neutrinos to show regions where observed limits are driven by under-fluctuations of data.

We have established the best direct-detection limits on dark matter-electron scattering in the mass range of \SIrange{16}{56}{\MeV\per\c\squared} for a heavy mediator and above \SI{80}{\MeV\per\c\squared} for a light mediator, with newly excluded parameter space shaded red.
These new \DSfs results on \LDM-electron scatters improve upon those previously obtained in 2018~\cite{DS50WimpER2018} primarily due to the refined data selection criterion which suppresses correlated events between \SI{4}{\el} and \SI{7}{\el}~\cite{DSLongPaper}.
Additional sensitivity gain comes from improved data selection, a more accurate detector calibration, improved background modeling, and a larger data-set.

We have also placed the first constraints on galactic axion-like particles and dark photons with an argon target.
Stronger direct-detection limits are placed on both \AxionElecCoupling and \DarkPhotonKinMixing for masses between \SI{0.03}{\keV\per\c\squared} and \SI{0.2}{\keV\per\c\squared}.

\DSfs is the first \DM direct-detection experiment to set limits on the sterile neutrino mixing angle \SnMixingAngle. 
Under the Standard Halo Model assumption, our results improve upon existing direct limits set by a high-precision measurement of the $^{63}$Ni~$\beta$ spectrum~\cite{HolzschuhSearchHeavyNeutrinos1999}. 
However, these are well above the indirect detection limits set by the NuSTAR experiment~\cite{RoachNuSTARTestsSterileneutrino2020}, which looks for anomalous X-ray lines from radiative sterile neutrino \DM decays.

The upcoming DarkSide-20k experiment has a planned exposure almost four orders of magnitude larger than \DSfs and will provide more sensitive searches for each \DM model considered here.

\begin{acknowledgements}
The DarkSide Collaboration offers its profound gratitude to the LNGS and its staff for their invaluable technical and logistical support. We also thank the Fermilab Particle Physics, Scientific, and Core Computing Divisions. Construction and operation of the DarkSide-50 detector was supported by the U.S. National Science Foundation (NSF) (Grants No. PHY-0919363, No. PHY-1004072, No. PHY-1004054, No. PHY-1242585, No. PHY-1314483, No. PHY-1314501, No. PHY-1314507, No. PHY-1352795, No. PHY-1622415, and associated collaborative grants No. PHY-1211308 and No. PHY-1455351), the Italian Istituto Nazionale di Fisica Nucleare, the U.S. Department of Energy (Contracts No. DE-FG02-91ER40671, No. DEAC02-07CH11359, and No. DE-AC05-76RL01830),  the Polish NCN (Grant No. UMO-2019/33/B/ST2/02884) and the Polish Ministry for Education and Science (Grant No. 6811/IA/SP/2018). We also acknowledge financial support from the French Institut National de Physique Nucl\'eaire et de Physique des Particules (IN2P3),   the  IN2P3-COPIN consortium (Grant No. 20-152),  and the UnivEarthS LabEx program (Grants No. ANR-10-LABX-0023 and No. ANR-18-IDEX-0001),  from the São Paulo Research Foundation (FAPESP) (Grant No. 2016/09084-0),  from the Interdisciplinary Scientific and Educational School of Moscow University ``Fundamental and Applied Space Research'',  from the Program of the Ministry of Education and Science of the  Russian  Federation  for  higher  education  establishments,  project No. FZWG-2020-0032 (2019-1569),  from IRAP AstroCeNT funded by FNP from ERDF, and from the Science and Technology Facilities Council, United Kingdom.  I.~Albuquerque is partially supported by the Brazilian Research Council (CNPq). This project has received funding from the European Union’s Horizon 2020 research and innovation program under grant agreement No 952480. Isotopes used in this research were supplied by the United States Department of Energy Office of Science by the Isotope Program in the Office of Nuclear Physics.
 \end{acknowledgements}

\bibliography{biblio}
\end{document}